\documentclass[a4paper,11pt]{article}
\usepackage{pos}
\pdfoutput=1

\usepackage{graphicx}
\usepackage{xcolor}
\usepackage{booktabs}
\usepackage{caption}
\usepackage{xspace}
\usepackage{siunitx}
\usepackage[export]{adjustbox}
\newlength\fw
\newlength\fh
\begin{document}

\title{MRHS multigrid solver for Wilson-clover fermions}

\ShortTitle{MRHS multigrid solver for Wilson-clover fermions}

\author{Daniel Richtmann}
\author*{Nils Meyer}
\author{Tilo Wettig}

\affiliation{Department of Physics, University of Regensburg, 93040 Regensburg, Germany}

\emailAdd{nils.meyer@ur.de}

\abstract{%
  We describe our implementation of a multigrid solver for Wilson-clover fermions, which increases parallelism by solving for multiple right-hand sides (MRHS) simultaneously.
  The solver is based on Grid and thus runs on all computing architectures supported by the Grid framework.
  We present detailed benchmarks of the relevant kernels, such as hopping and clover term on the various multigrid levels, intergrid operators, and reductions.
  The benchmarks were performed on the JUWELS Booster system at Jülich Supercomputing Centre, which is based on Nvidia A100 GPUs.
  For example, solving a $24^3\times128$ lattice on 16 GPUs, the overall speedup obtained solely from MRHS is about 10x.
}

\FullConference{%
  The 39th International Symposium on Lattice Field Theory, LATTICE2022 8th-13th August, 2022\\
  University of Bonn, Germany
}

\maketitle

\section{Introduction}

The numerically dominant operation in Lattice QCD is the inversion of the Dirac matrix.
As we approach the continuum limit and the limit of physical quark masses, the condition number of this matrix increases dramatically.
The inversion then requires sophisticated algorithms, in particular multigrid (MG) methods and variants thereof \cite{Luscher:2003qa,Luscher:2007se,Brannick:2007ue,Babich:2010qb,Frommer:2013fsa}.
Multigrid methods employ a hierarchy of fine and coarse grids.
This creates some issues when trying to exploit modern hardware, which offers ever increasing degrees of parallelism.
The small local volume on coarser grids cripples the performance on GPUs already now.
Our approach to address this issue is to reformulate the problem such that a large number of degrees of freedom are processed in parallel, i.e., we solve multiple right-hand sides (MRHS) simultaneously \cite{Morgan:2004zh,Richtmann:2016kcq}.

The structure of this paper is as follows.
In Sec.~\ref{sec:mrhs} we describe the motivation for and the implementation of our MRHS solver.
In Sec.~\ref{sec:booster} we briefly review the architecture of the JUWELS Booster machine on which our benchmarks were executed.
In Secs.~\ref{sec:fine} through \ref{sec:inner} we present a number of benchmarks for important elements of the solver: hopping and clover term on the fine grid, intergrid operators, hopping term on the coarse grids, and inner products.
In Sec.~\ref{sec:cls} we present the overall performance of the solver for a particular CLS configuration and break down the various contributions to the wall-clock time.
We conclude in Sec.~\ref{sec:summary}.
The MRHS code presented in this paper is publicly available at \url{https://github.com/DanielRichtmann/Grid}.

\section{Multiple right-hand sides: Motivation and Grid implementation}
\label{sec:mrhs}

There are two main benefits of MRHS: (i) Since the gauge fields can be reused if the Dirac matrix is inverted for MRHS we obtain a higher cache reuse, which leads to improved compute performance.
(ii) With MRHS we have larger and fewer messages in communication and thus move from a latency problem to a bandwidth problem.

We use Grid \cite{Boyle:2015tjk} for our implementation.
Grid supports native 5-dimensional fields and operations for domain-wall fermions (DWF).
DWF is implicitly an MRHS system, and therefore the groundwork for an MRHS formulation of Wilson fermions is already in place.
We formulate MRHS as a 5d problem with the RHS index in the fifth dimension \cite{Alappat:2021icl}.
This allows us to call the hopping-term kernel function used in DWF also for the MRHS Wilson-clover hopping term.
We implemented all other MG kernels using the same data layout and loop structure as in the hopping term.
Therefore no data-layout transformations are necessary inside the solver.

All benchmarks for individual components of the MG solver presented in this paper were performed in single precision on random gauge fields (cf.~\texttt{HotConfiguration}) and random sources.

\section{JUWELS Booster architecture}
\label{sec:booster}

\begin{figure}
  \begin{minipage}[t]{\linewidth}
    \begin{center}
      \includegraphics[width=0.65\linewidth]{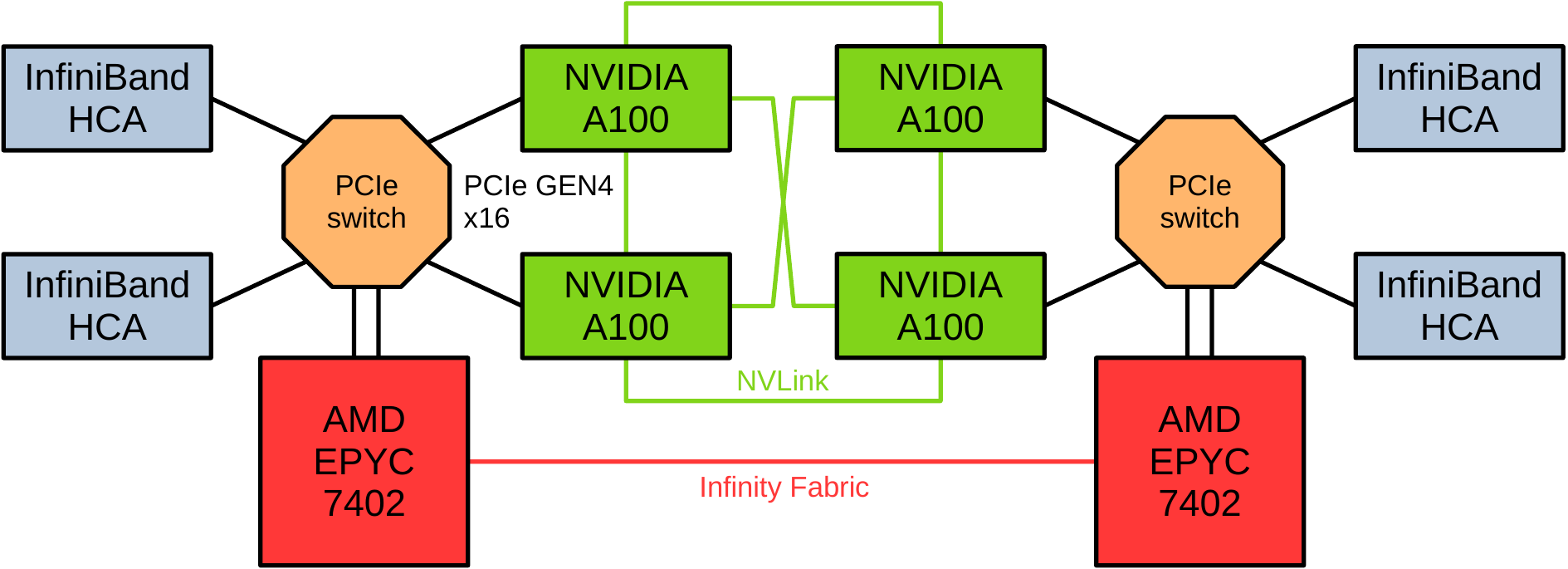}\\
      \caption{\label{fig:booster}Block diagram of the JUWELS Booster node architecture.}
    \end{center}
  \end{minipage}
\end{figure}

We performed benchmarks of our code on the JUWELS Booster machine \cite{JUWELS-Booster} at the Jülich Supercomputer Centre (JSC).
The Booster comprises 936 identical compute nodes.
A single node is shown in Fig.~\ref{fig:booster}.
It is equipped with dual-socket AMD EPYC 7402 processors, four NVIDIA A100 GPUs with 40~GB HBM2 memory each, interconnected via NVLink, and four InfiniBand HDR Host Channel Adapter (HCA) cards, each providing 200 Gbit/s per direction.
The network has a Dragonfly+ topology \cite{7885210}, with cells of 48 nodes each which are connected in a full fat tree.

\section{Hopping and clover term on the fine level}
\label{sec:fine}

\begin{figure}
  \centering
  \begin{tabular}{c@{\hspace*{10mm}}c}
  \includegraphics[height=\fh]{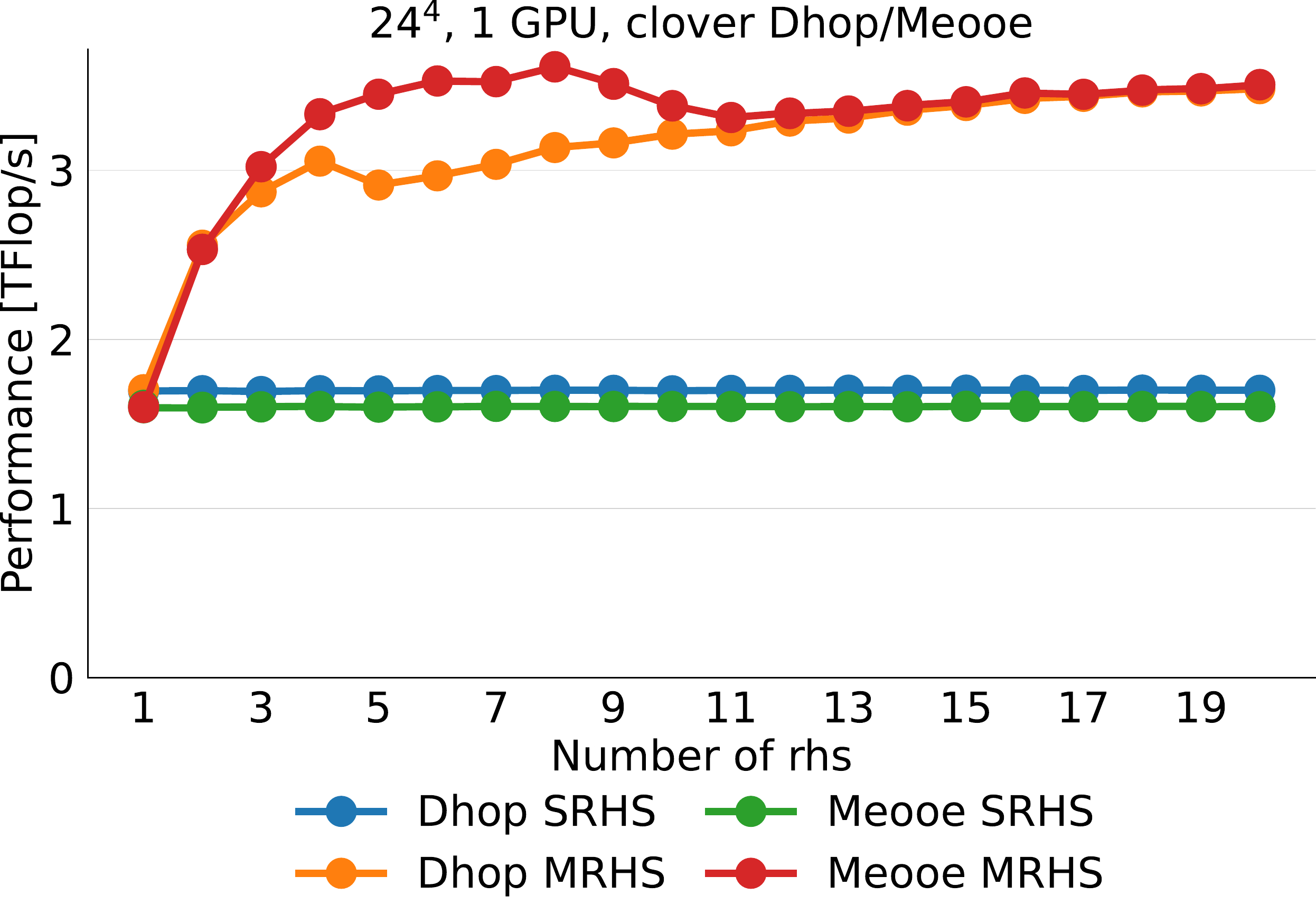}
  & \includegraphics[height=\fh]{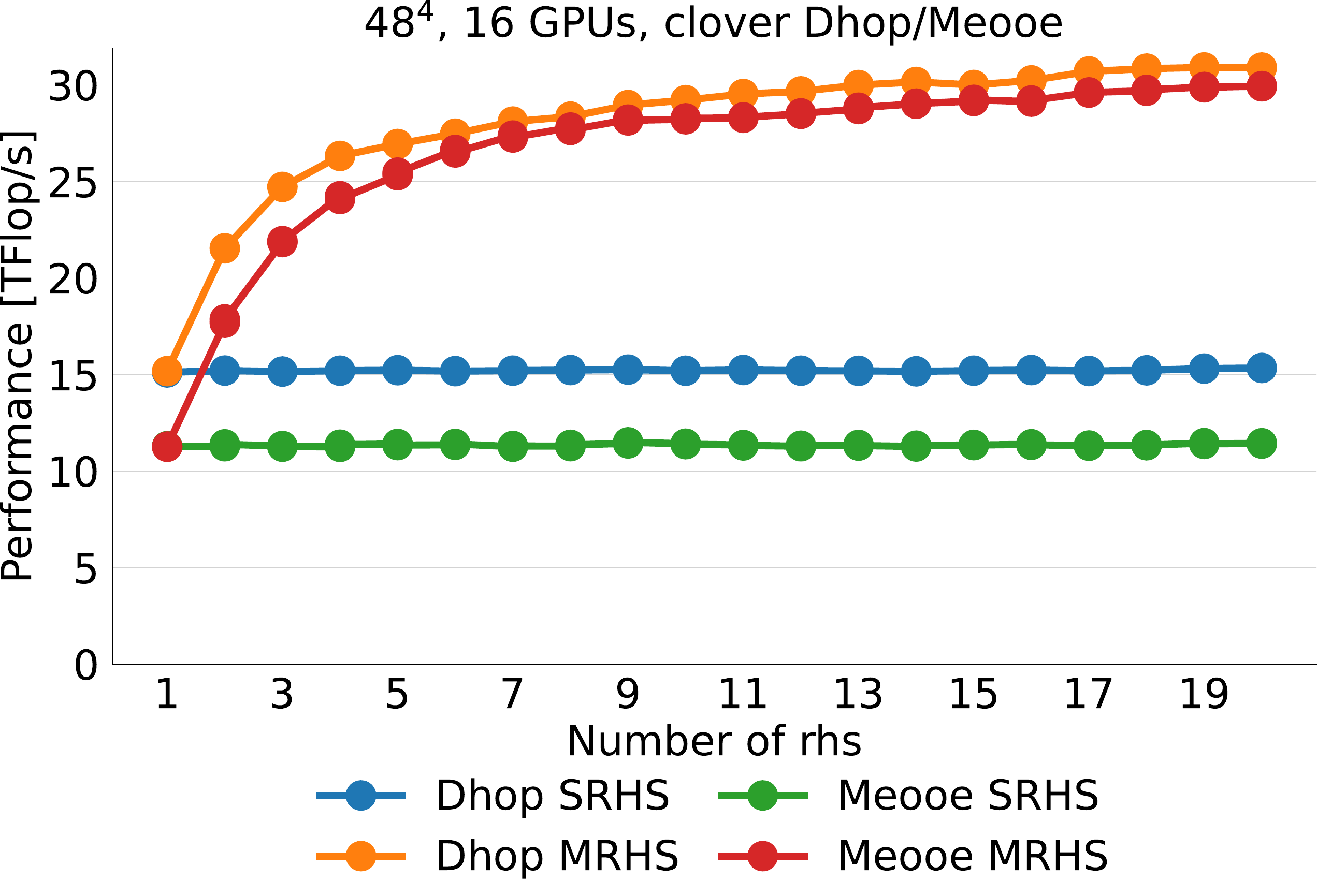}
  \end{tabular}
  \caption{%
    \label{fig:fine1}Performance of the hopping term as a function of the number of RHS on the fine level.
    The lattice volumes are $24^4$ on a single GPU (left) and $48^4$ on 16 GPUs = 4 nodes (right).
    SRHS stands for single right-hand side.
  }
\end{figure}

In Fig.~\ref{fig:fine1} we present benchmarks for the hopping term on the fine MG level.
In these plots, as well as in other plots below, $D_\text{hop}$ stands for the hopping term on the full lattice and $M_\text{eooe}$ for the hopping term between checkerboarded lattices.
Note that $D_\text{hop}$ is not implemented as two successive calls to $M_\text{eooe}$ on the even and odd lattices since $D_\text{hop}$ acts on the full lattice and thus the ordering of the sites is different in Grid.
Thus, depending on whether one works on the full lattice or on the checkerboarded lattices, one has to call $D_\text{hop}$ once or $M_\text{eooe}$ twice, respectively.

In Fig.~\ref{fig:fine1} we see that using MRHS for $D_\text{hop}$ gives a speedup of about 2x on both 1 GPU and 16 GPUs, while for $M_\text{eooe}$ the speedup is about 2.2x on 1 GPU and 2.6x on 16 GPUs.
The performance difference between $D_\text{hop}$ and $M_\text{eooe}$ is most likely due to the different site orderings (see above) and the difference in message sizes.

\begin{figure}
  \centering
  \includegraphics[height=\fh]{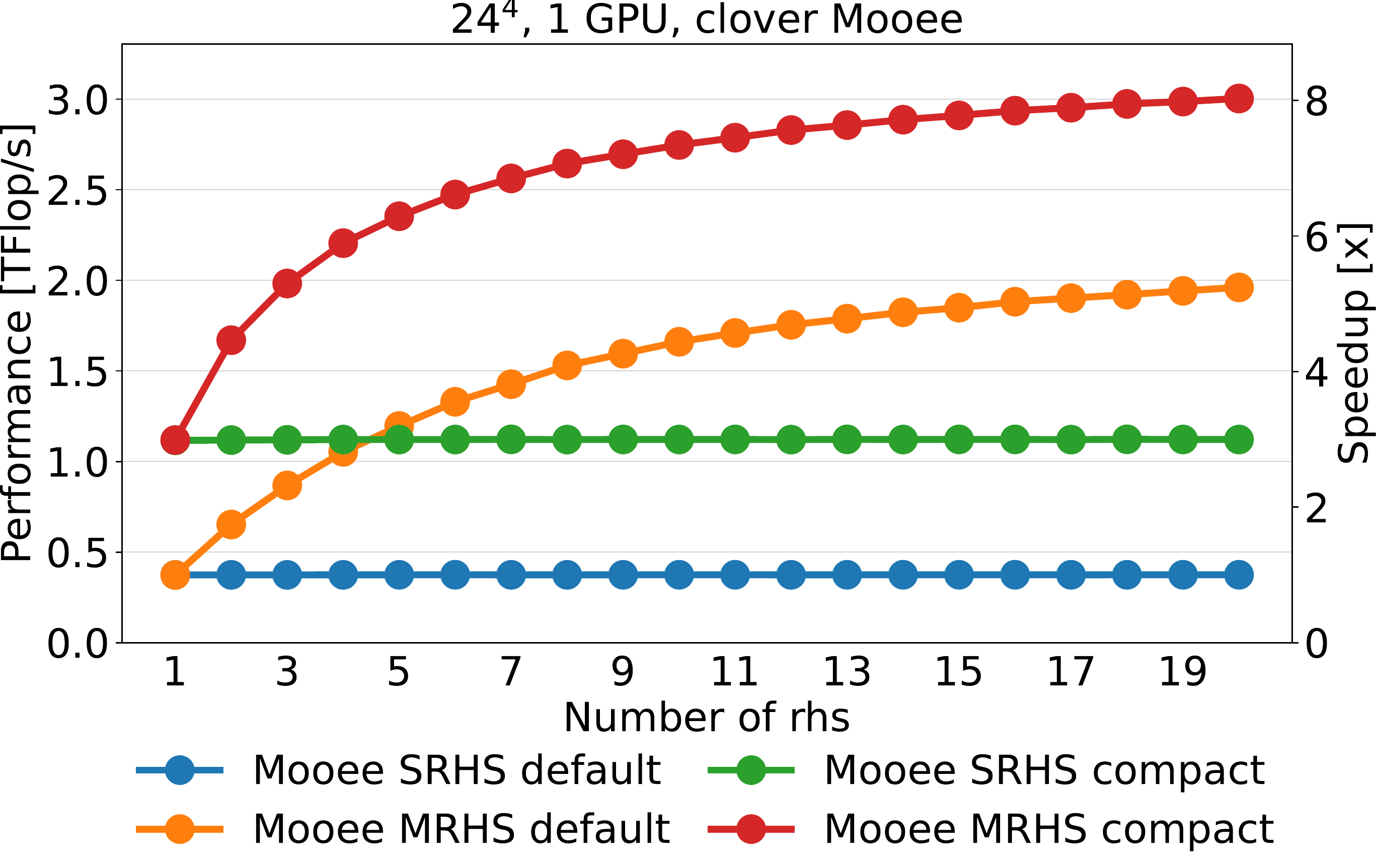}
  \caption{%
    \label{fig:fine2}Performance and speedup of the clover term as a function of the number of RHS on the fine level, on a single GPU with a lattice volume of $24^4$.
    The performance of applying the precomputed clover term scales trivially with the number of GPUs as this term is then diagonal with respect to lattice sites.
  }
\end{figure}

In Fig.~\ref{fig:fine2} we show the performance of the clover term, denoted by $M_\text{ooee}$.
Here, ``default'' stands for the default Wilson-clover class provided by Grid, while ``compact'' stands for our faster implementation of the clover term (\texttt{CompactWilsonCloverFermion}), which is also available in upstream Grid.
We see that MRHS leads to a speedup of about 5.5x for the default method and a combined speedup of about 8x for the improved compact method.

Note that there is a saturation effect in both figures.
As we can also see from the figures in the following sections, increasing the number of RHS beyond $20\!\sim\!30$ does not lead to a significant gain in performance anymore in most cases.

\section{Intergrid operators for three-level multigrid}
\label{sec:intergrid}

\begin{figure}
  \centering
  \begin{tabular}{c@{\hspace*{10mm}}c}
    {\small fine to intermediate} & {\small intermediate to coarse} \\[1mm]
    \includegraphics[height=\fh]{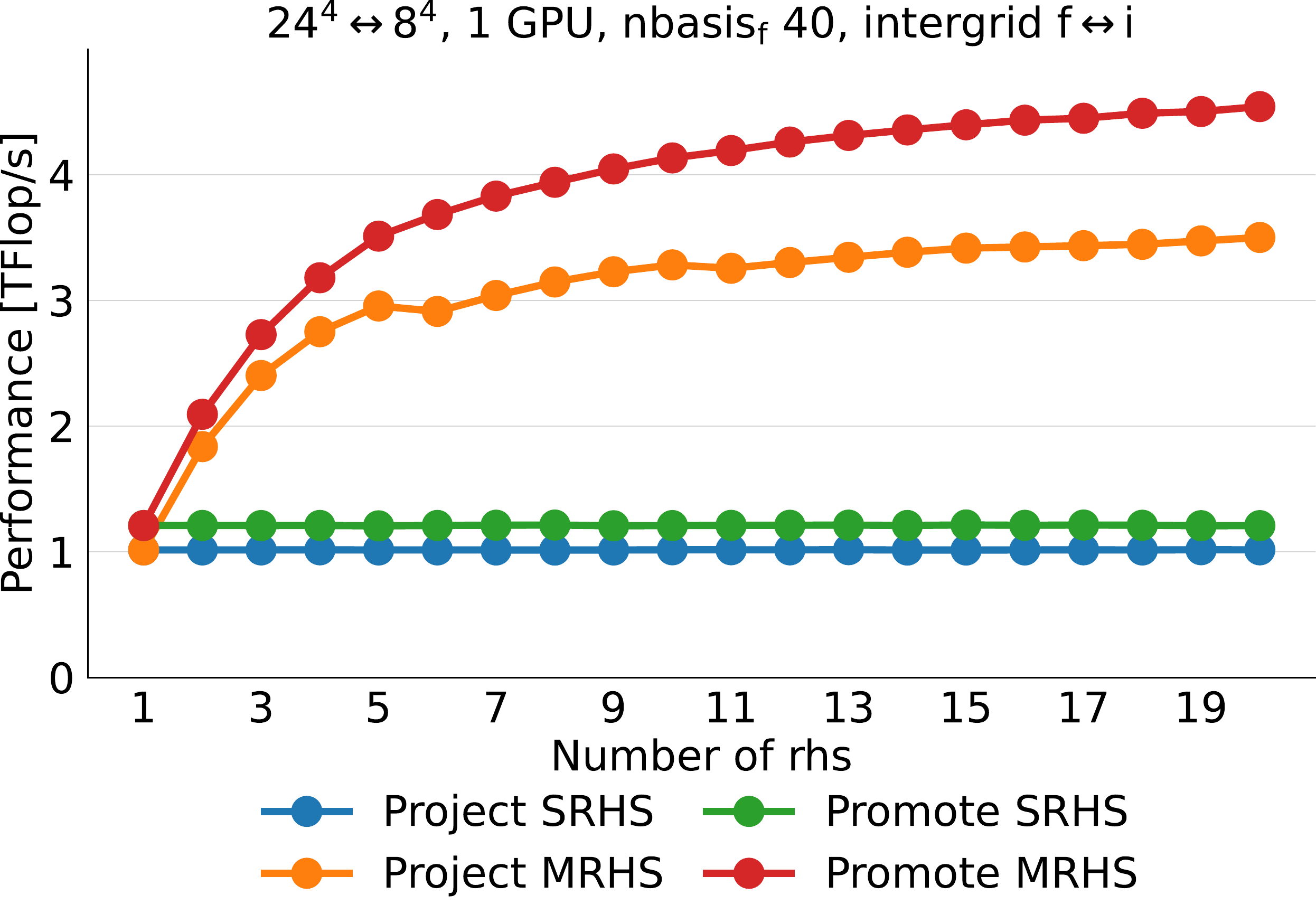}
    & \includegraphics[height=\fh]{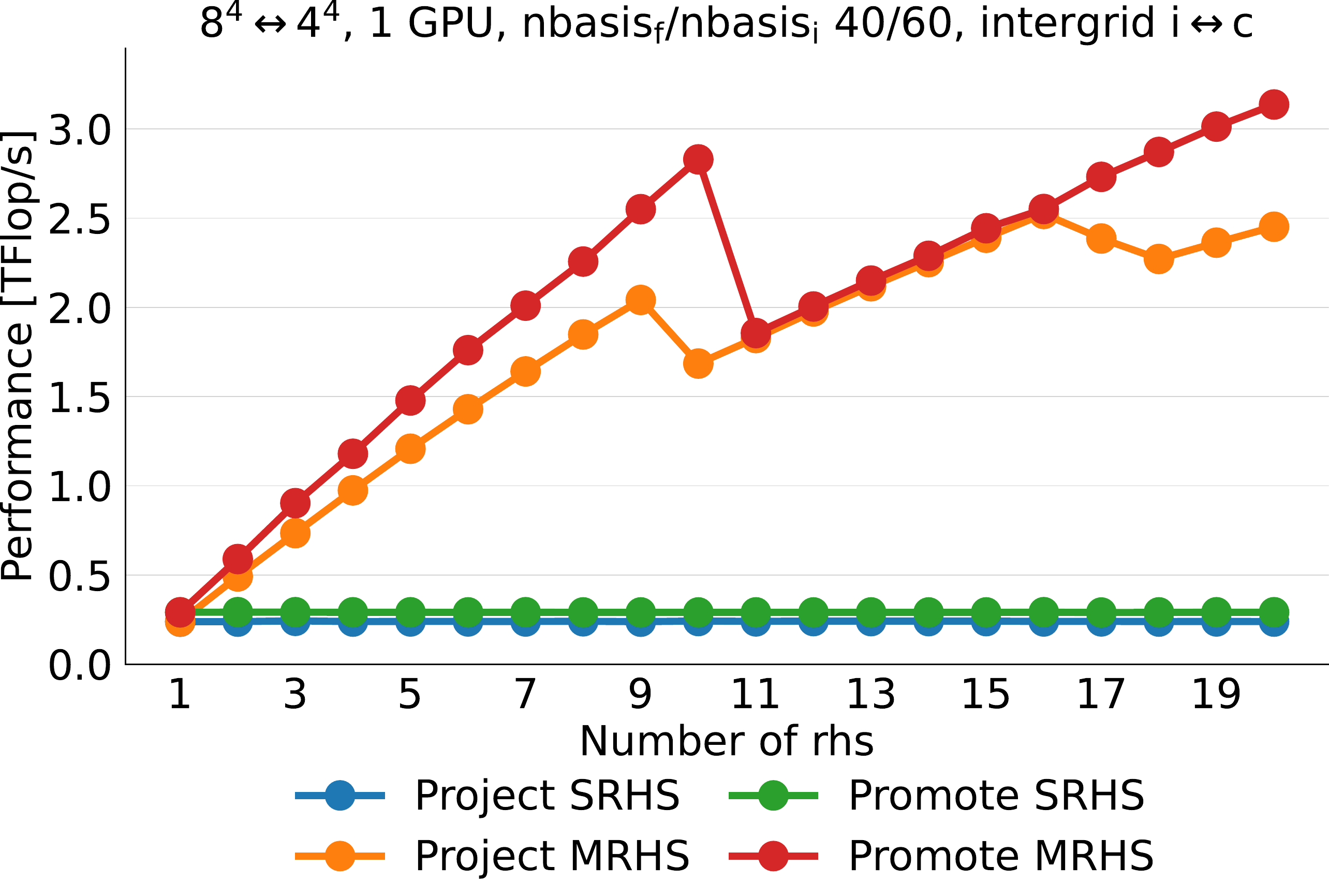}
  \end{tabular}
  \caption{%
    \label{fig:intergrid}Performance of intergrid operators as a function of the number of RHS.
    Here, the symbols f, i, and c stand for the fine, intermediate, and coarse level, respectively.
  }
\end{figure}

In Fig.~\ref{fig:intergrid} we present, for a three-level multigrid, benchmarks of the intergrid operators that are needed to project the fields from the finer to the coarser levels and to promote the fields from the coarser to the finer levels.
Here, nbasis stands for twice the number of multigrid basis vectors.
The subscript on nbasis indicates the multigrid level.
Note that nbasis on a given level determines the size of the projection operators between this level and the next coarser level and equals the number of spinor elements per site at the next coarser level.
For example, in Fig.~\ref{fig:intergrid} we chose nbasis$_\text{f}=40$ and nbasis$_\text{i}=60$.
Therefore the spinors on the intermediate grid have 40 elements, the projection operator from intermediate to coarse level has dimension $60\times40$ per aggregate, and the spinors on the coarse grid have 60 elements.

For the intergrid operators between fine and intermediate level we see that using MRHS gives a speedup of about 3.5x, while for the operators between intermediate and coarse level the speedup of about 10x is much larger.
This is caused by the very small local volume on the coarsest grid and thus a severe underutilization of the GPUs for SRHS.
In both plots the performance is higher for the promotion compared to the projection since for the promotion we can parallelize over the finer lattice, while for the projection we parallelize over the coarser lattice to avoid synchronizations that would otherwise be necessary for parallel reduction within each aggregate.
Thus more parallelism is available for the promotion.

The lack of saturation in the intermediate-to-coarse plot indicates that higher speedups could be obtained by going to a larger number of RHS.
We did not investigate this further since the contribution of the intergrid operators to the overall wall-clock time is very small, see Fig.~\ref{fig:U101}.

\section{\boldmath Coarse-level operators $D_c$}
\label{sec:coarse}

\begin{figure}
  \centering
  \begin{tabular}{lr}
    \includegraphics[height=\fh]{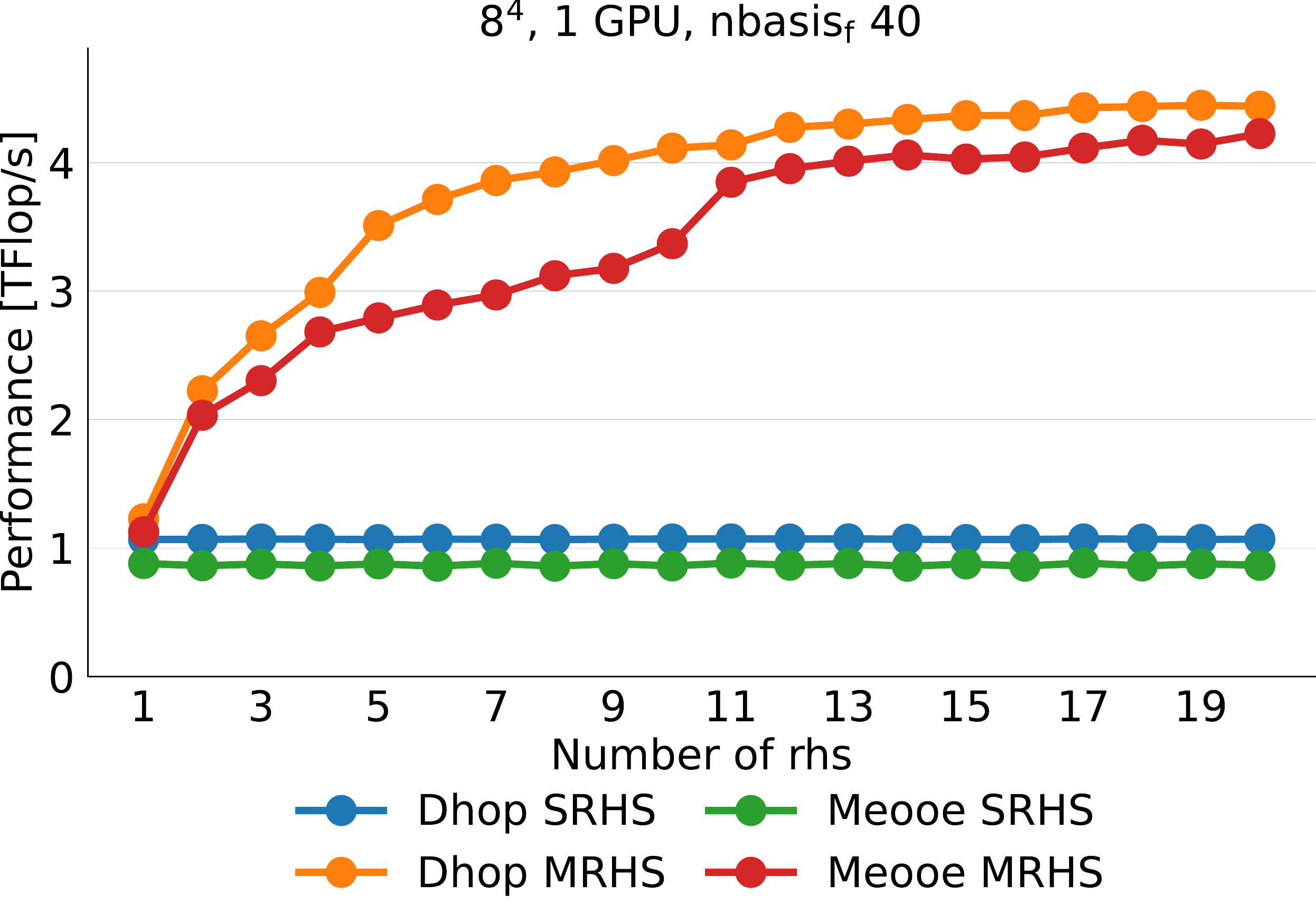}
    & \includegraphics[height=\fh]{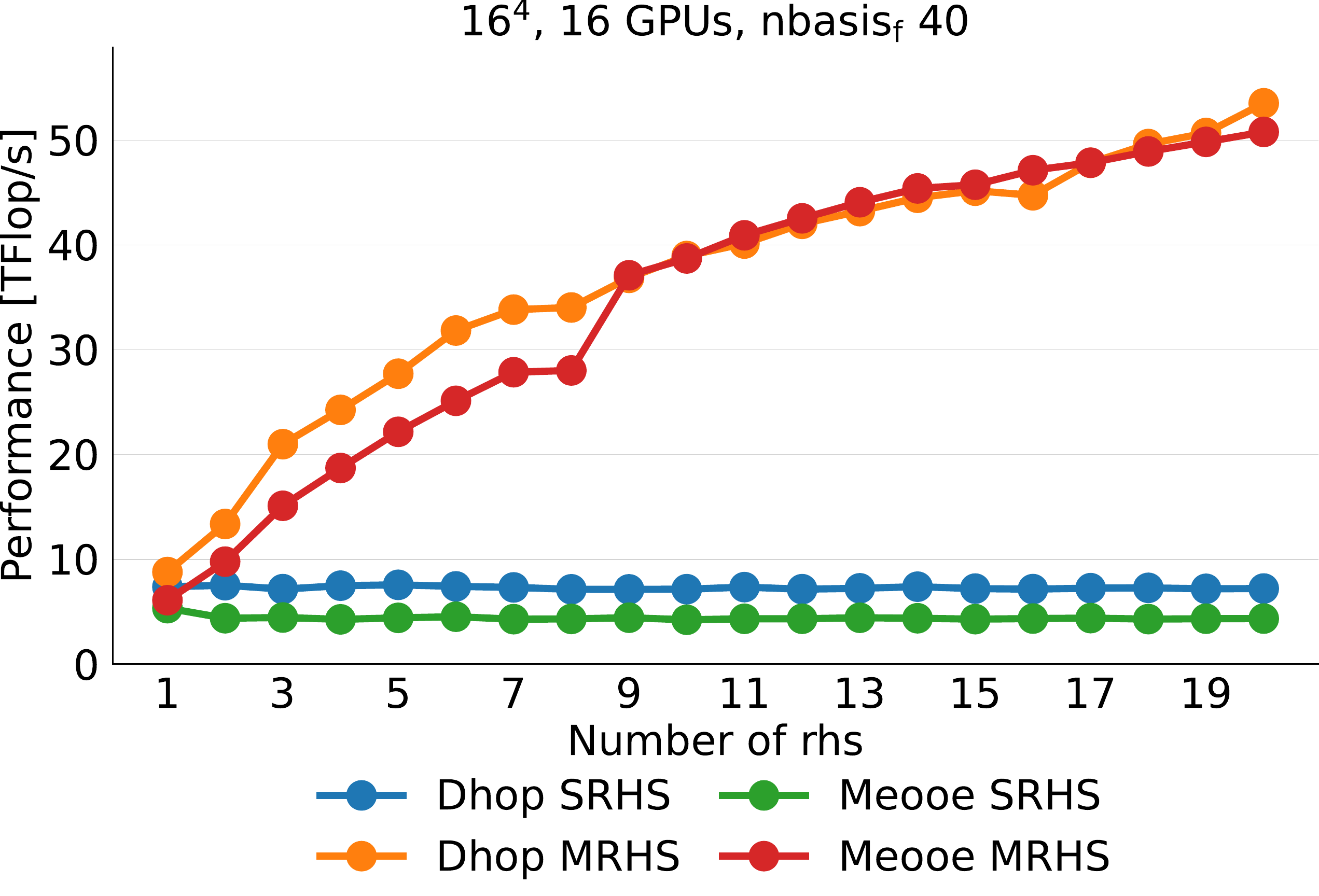}\\
    \includegraphics[height=\fh]{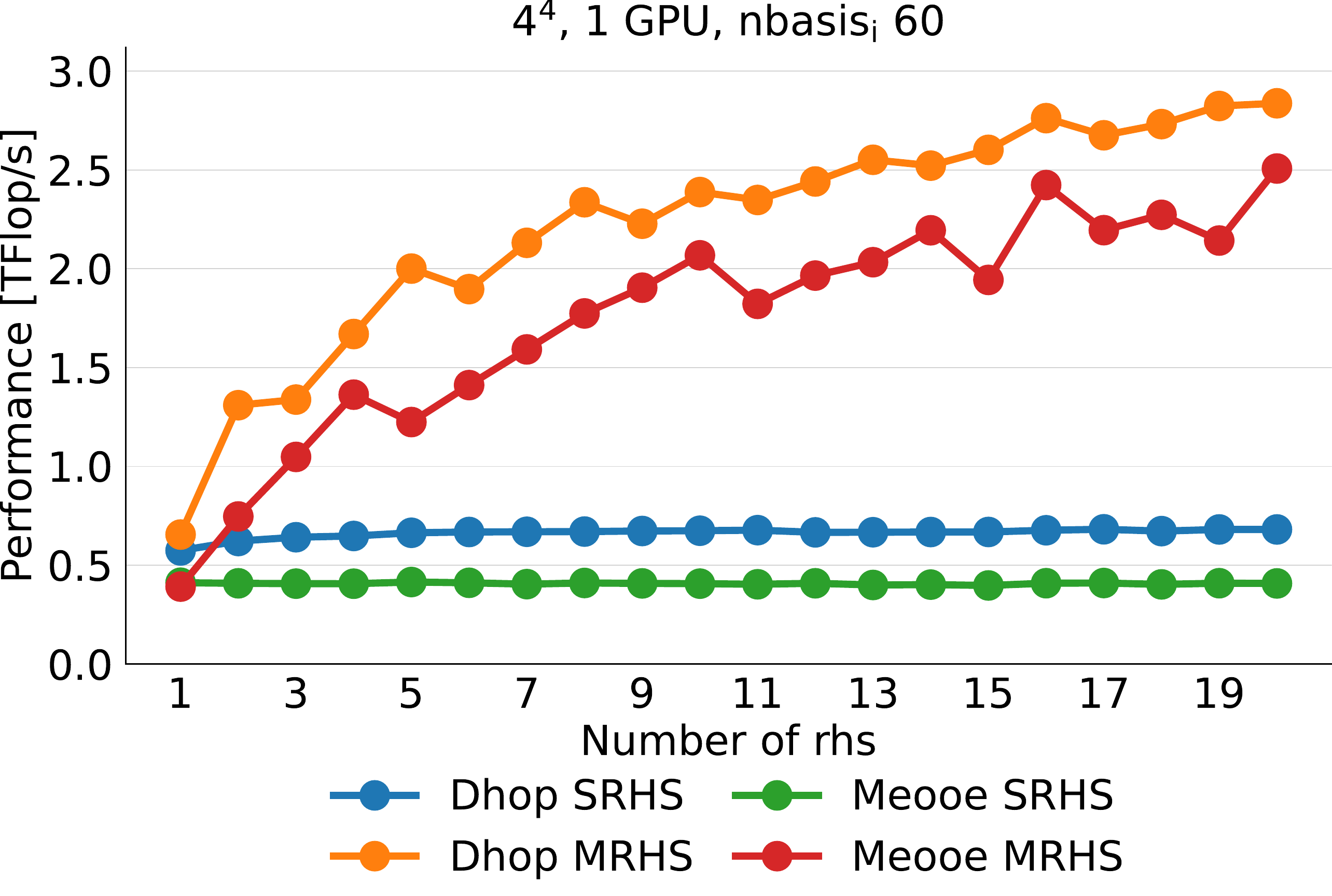}
    & \includegraphics[height=\fh]{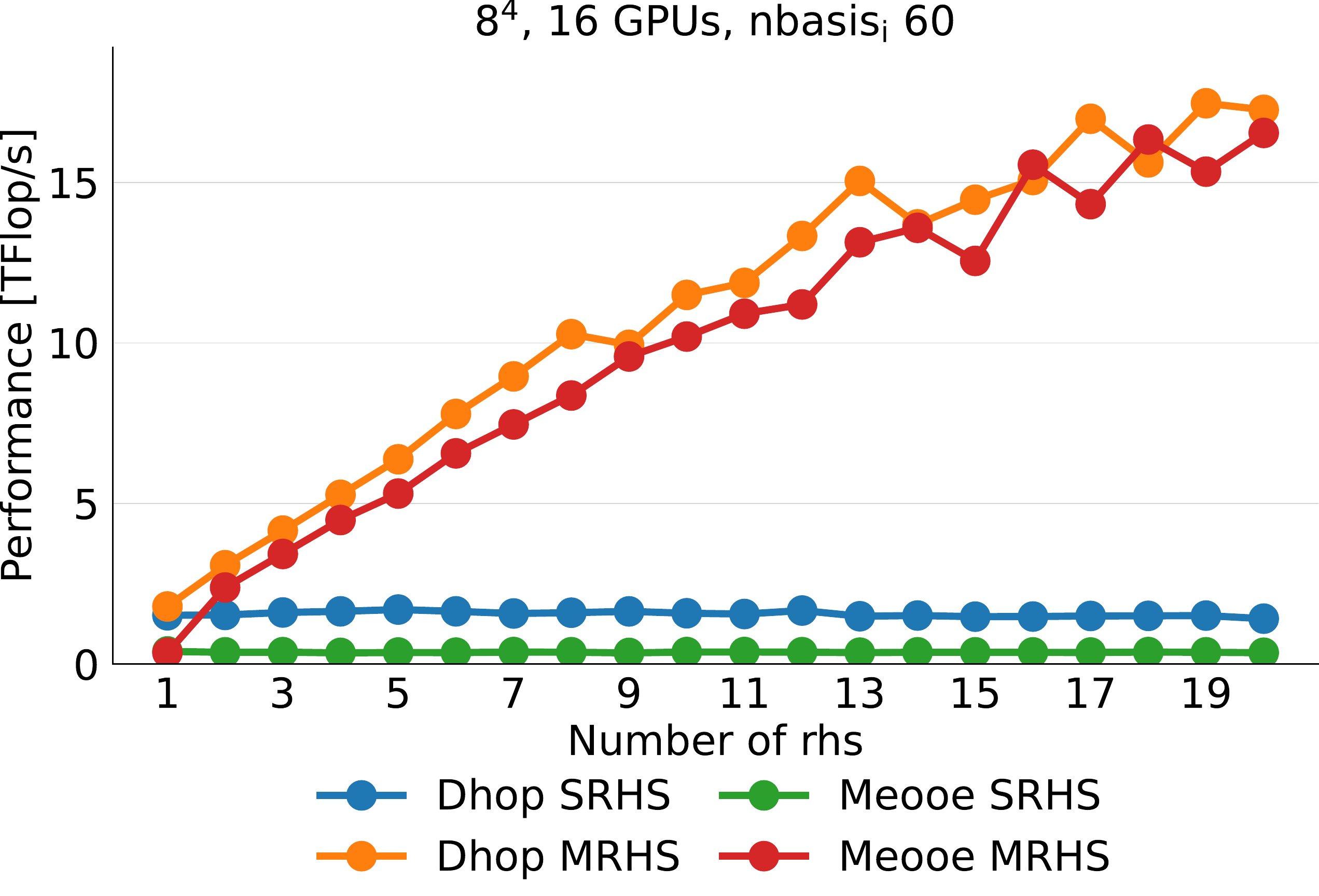}
  \end{tabular}
  \caption{%
    \label{fig:coarser}Performance of the hopping term on coarser levels: intermediate level (top) and coarse level (bottom) on 1 GPU (left) and 16 GPUs (right).
  }
\end{figure}

In Fig.~\ref{fig:coarser} we show the performance of the hopping term on the coarser levels for a three-level multigrid.
(We do not show the clover terms on the coarser levels since their contribution to the run time is negligible.) On the intermediate level we observe speedups of about 4x ($D_\text{hop}$) and 5x ($M_\text{eooe}$) for a single GPU and about 7x ($D_\text{hop}$) and 12x ($M_\text{eooe}$) for 16 GPUs.
On the coarsest level the speedup increases to about 4x ($D_\text{hop}$) and 7x ($M_\text{eooe}$) for a single CPU and to about 12x ($D_\text{hop}$) and 47x ($M_\text{eooe}$) for 16 GPUs.
Furthermore, saturation is not reached yet on the coarsest level.
Clearly, MRHS leads to larger speedups on coarser levels, or, equivalently, smaller volumes, which is again due to the severe lack of parallelism on the coarser levels without using MRHS.
For the same reason, $M_\text{eooe}$ performs worse than $D_\text{hop}$ since it acts on half the data.

\section{Reductions: Inner product of two vectors}
\label{sec:inner}

\begin{figure}
  \centering \includegraphics[height=.95\fh]{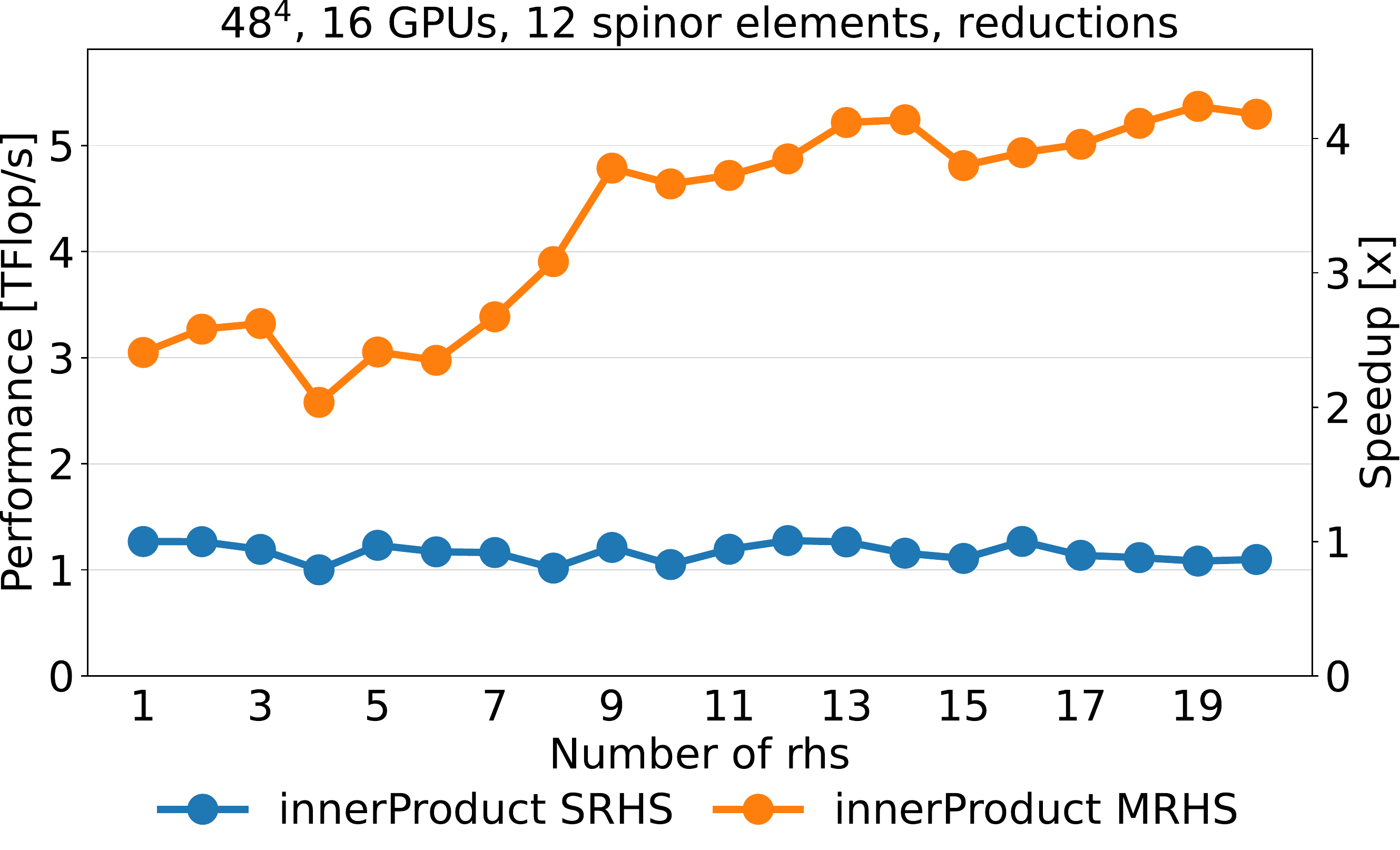} \hfill \includegraphics[height=.95\fh]{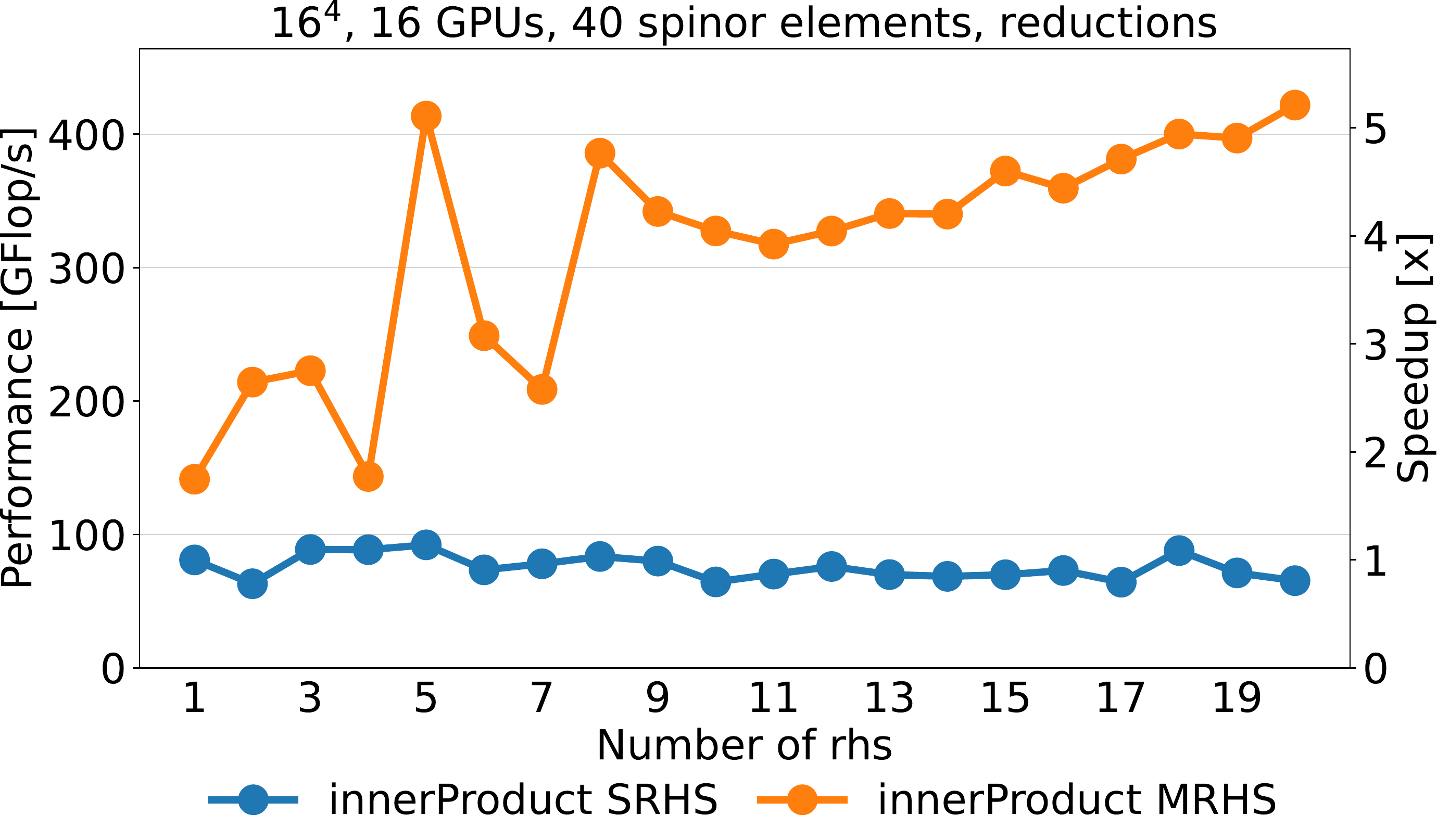}\\[6mm]
  \includegraphics[height=.95\fh]{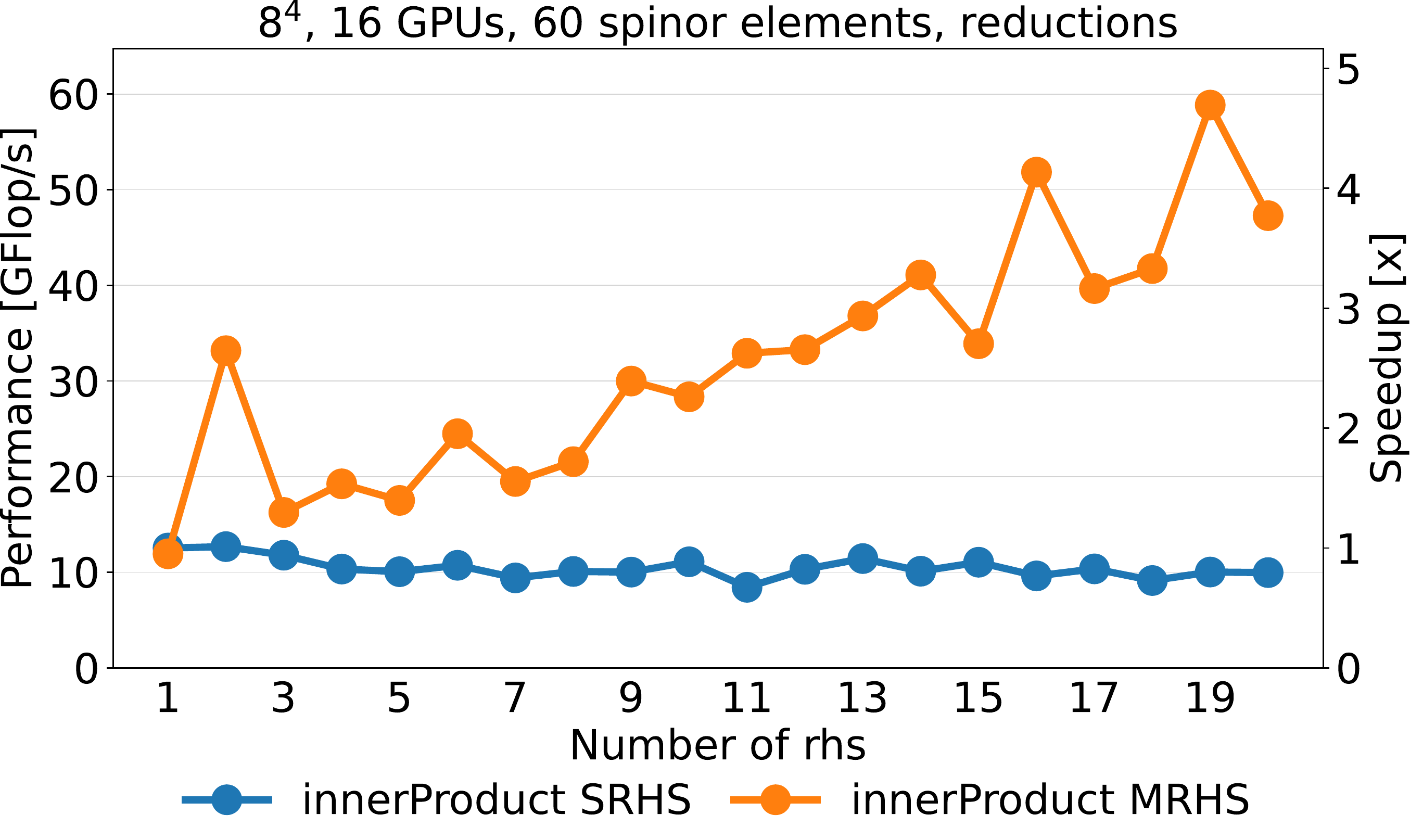}
  \caption{%
    \label{fig:reductions}Performance and speedup for sequential and simultaneous computation of inner products on the fine (top left), intermediate (top right), and coarsest (bottom) level.
  }
\end{figure}

As an example of reductions we show in Fig.~\ref{fig:reductions} the performance of the inner product of two vectors as a function of the number of inner products.
The three plots are for the three MG levels, all on 16 GPUs.
Furthermore, ``spinor elements'' stands for the number of elements of the vector per lattice site (see the discussion in Sec.~\ref{sec:intergrid}), which is fixed at compile time.
The speedup is about 4-5x on all three levels.
However, note that the performance drops by roughly a factor of 10 when going from a given level to the next coarsest level.
In our benchmarks we observed that inner products contribute significantly to the run time on all multigrid levels if MRHS are used since the matrix part is sped up by a much larger factor than the inner product part.
Therefore further optimizations of inner products would certainly have an impact.

\section{Overall speedup of two-level multigrid on CLS configuration U101}
\label{sec:cls}

In this section we present the performance of a two-level multigrid method on a real-world example, i.e., configuration U101 of the CLS collaboration.
The parameters of this configuration are as follows: $V=24^3\times128$, $N_f = 2+1$, $L\sim\SI{2}{fm}$, $m_\pi = \SI{280}{MeV}$, $m_K = \SI{467}{MeV}$.

\begin{figure}
  \centering
  \parbox{.6\textwidth}{\includegraphics[height=\fh]{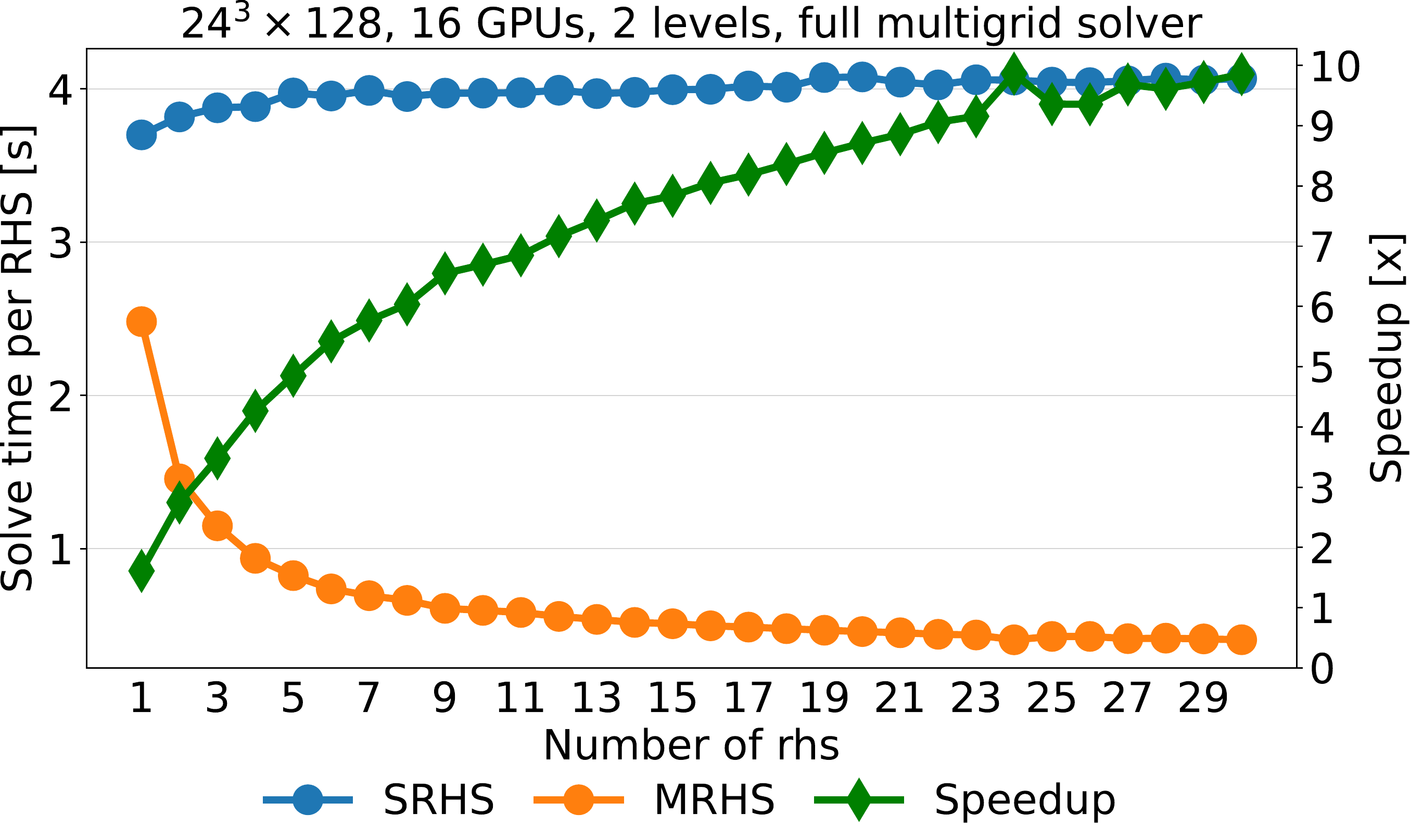}}
  {\small\begin{tabular}{lr}
    Fine                 &  4.78 \\
    Project              &  0.03 \\
    Promote              &  0.02 \\
    Coarse               &  6.29 \\
    Rest                 &  0.23 \\
    Total preconditioner & 11.35 \\
    Full solver          & 12.16
  \end{tabular}}
  \caption{%
    \label{fig:U101}Left: Solve time per RHS of an FGMRES preconditioned by a two-level multigrid method for CLS configuration U101 on 16 GPUs.
    SRHS refers to sequential solves for single RHS, while MRHS refers to a simultaneous solve for multiple RHS.
    Right: Breakdown of the contributions to the MRHS solve time for 30 RHS, given in seconds.
  }
\end{figure}

Figure~\ref{fig:U101} shows the solve time and the speedup of the full solver as a function of the number of RHS.
The full solver is an FGMRES preconditioned by the two-level multigrid method.
We approach saturation at about 30 RHS and obtain a speedup of about 10x.
We also break down the individual contributions of the various building blocks to the wall-clock time.
(The contributions of the inner products are in ``Fine'' and ``Coarse.'')
We observe that the intergrid operators have a negligible impact on performance and that most of the time is spent on the coarse level.
We also note that the preconditioner makes up for over 93\% of the overall solver wall-clock time.

\section{Summary and outlook}
\label{sec:summary}

Modern hardware forces us to exploit more and more levels of parallelism.
We have shown that our MRHS implementation of the multigrid preconditioner in Grid increases parallelism straightforwardly, which leads to a significant boost in performance.
The lesson from this example is obvious: when developing new or improving existing algorithms, one should try to find versions that can use multiple right-hand sides.

There are several directions for further improvements.
First, we will
make more optimization efforts on inner products.
Second, we will implement an eigenvector-recycling solver (GCRO-DR) on the coarsest level.
Finally, we will use Schwarz preconditioning on all MG levels to avoid communication.
Work in these directions is in progress.

\section*{Acknowledgments}

This work was supported by Deutsche Forschungsgemeinschaft through the PUNCH4NFDI consortium (DFG fund NFDI 39/1) and by the Competence Network for Scientific High Performance Computing (KONWIHR) funded by the Bavarian State Ministry of Science and the Arts.
We thank Christoph Lehner for stimulating discussions.

\bibliographystyle{JHEP_lat22}
\bibliography{references}

\end{document}